\documentclass[12pt]{iopart}
\usepackage{graphicx}
\usepackage{cite}
\usepackage{iopams}  
\usepackage{setstack}
\usepackage{bm}

\begin{document}

\title[Quantum correlations in a two-atom system]{Evolution of quantum
  correlations in a two-atom system}

\author{Ryszard Tana\'s}

\address{Nonlinear Optics Division, Faculty of Physics, Adam
  Mickiewicz University, Umultowska 85, 61-614 Pozna\'n, Poland}

\ead{tanas@kielich.amu.edu.pl}
\begin{abstract}
  We discuss the evolution of quantum correlations for a system of two
  two-level atoms interacting with a common reservoir.  The Markovian
  master equation is used to describe the 
  evolution of various measures of quantum correlations. It is shown
  that different measures of quantum correlations exhibit
  qualitatively different behaviour in their evolution. 
\end{abstract}



 \maketitle


\section{Introduction}
It is now well established that quantum entanglement, which is a
measure of quantum correlations, is a necessary resource to perform
some quantum information algorithms~\cite{horodecki09:_quant}. 
However, it has been realized that quantum entanglement does not
include all quantum correlations, and there are separable states which
exhibit quantum correlations different from entanglement. 
It became important to recognise the nature of correlations and
properly discriminate quantum correlations from the classical ones. A
number of measures of quantum correlations has been introduced, and
the most popular of them is quantum
discord~\cite{ollivier02:_quant,henderson01:_class} (see also the
review~\cite{modi11:_quant} and papers cited there). It is crucial,
from the point of view of future application, to know how quantum
correlations evolve in time when a multipartite system interacts with the
dissipative environment.

The time evolution of entanglement for a
system of two qubits, or two two-level atoms, can be qualitatively
described for various physical situations, and it has been studied
extensively in recent
years~\cite{zyczkowski01:_dynam,jakobczyk02:_entan,ficek03:_entan,nicolosi04:_dissip,yu04:_finit,almeida07:_envir,ficek06:_dark,ficek08:_delay}.
A lot of discussion has been devoted to the problem of disentanglement
of the two-qubit system in a finite time, despite the fact that all
the matrix elements of the two-atom system decay only
asymptotically. Yu and Eberly~\cite{yu04:_finit} coined the name
``entanglement sudden death'' (ESD) to the process of finite-time
disentanglement. ESD has recently been confirmed
experimentally~\cite{almeida07:_envir}.  Another problem related to
the entanglement evolution that attracted attention is the evolution
of the entangled qubits interacting with the non-Markovian
reservoirs~\cite{bellomo07:_non_markov,cao08:_non_markov,wang08:_decoh_markov}.
It has also been shown that squeezed reservoir leads to the
steady-state entanglement~\cite{tanas04:_station} and revivals of
entanglement~\cite{mundarain07:_decoh}.

Beside entanglement, a considerable amount of attention  has
been paid in recent years to quantum discord. Generally, quantum
discord is difficult to calculate because it requires the optimisation
procedure which usually amounts to extensive numerical
calculations. In case of two qubits analytical results can be obtained
for some specific families of states, like Bell-diagonal states or
states having maximally-mixed marginals~\cite{luo08:_quant}. Another
interesting family of states is a family of the two-qubit X-states,
which are of interest here. Ali~{\em et
  al.}~\cite{ali10:_quant_x} reported a closed form
solution for quantum discord of the X states. However, it turned out
that their algorithm is not universal. Lu~{\em et
  al.}~\cite{lu11:_optim} have proven that the universal set of
orthogonal projective measurements cannot be found for the full family
of X states. Some counterexamples have been given
in~\cite{lu11:_optim,chen11:_quant_x}. Chen~{\em et
  al.}~\cite{chen11:_quant_x}, however, confirmed applicability of the
algorithm for several special cases of X states. Lu~{\em et al.} have
found that the probability distribution of measurements is centralised
around a specific von Neumann measurement, which they called the
maximal-correlation-direction measurement. This observation justifies
the Ali~{\em et al.} algorithm in a statistical sense, because, it
turns out that for 99.4\% of cases of numerically generated X states the
algorithm gives correct results. The situation with arbitrary
two-qubit states is more complicated and the best what has been
achieved so far are the two transcendental equations obtained by
Girolami and Adesso~\cite{girolami11:_quant} which must be solved
numerically to get quantum discord. 

Daki\'c~{\em et al.}~\cite{dakic10:_neces} have introduced another
measure of quantum correlation based on the Hilbert-Schmidt distance
measure, which is called geometric discord. The advantage of this
measure is that it allows for getting analytical formulas for general
two-qubit states. Girolami and Adesso~\cite{girolami12:_obser}
introduced the so called observable measure of bipartite quantum
correlation which is a lower bound to the geometric discord. Behaviour
of the geometric discord under decoherence has been studied
in~\cite{lu10:_geomet}. Recently Bellomo {\it et al.}~\cite{bellomo12:_dynam} compared the
dynamics of geometric and entropic quantifiers of the different kinds
of correlations in a non-Markovian open two-qubit system under local
dephasing.

In this paper we study the evolution of concurrence (a measure of
entanglement), quantum discord and geometric discord in a system of
two two-level atoms interacting with a 
common reservoir being in a vacuum state. The evolution of the system is
described by the Markovian master equation introduced by
Lehmberg~\cite{lehmberg70:_radiat_n} and
Agarwal~\cite{agarwal74:_quant_statis_theor_spont_emiss_relat_approac},
taking into account the cooperative behaviour of the atoms. It is
shown that different measures of quantum correlations evolve in time
quite differently. We compare their evolution for a family of pure initial
states. Since the evolution of the system is described by a realistic
master equation, which is a good testing ground for studying physical
processes involving two atoms, we believe the results obtained shed
new light on understanding quantum correlations.  

\section{Master equation}
We consider a system of two two-level atoms, $A$ and $B$, with ground states
$|g_{i}\rangle$ and excited states $|e_{i}\rangle$ ($i=A,B$) connected
by dipole transition moments $\bm{\mu}_{i}$. The atoms are located at
fixed positions $r_{A}$ and $r_{B}$ and coupled to all modes of the
electromagnetic field being in the vacuum state.

The reduced two-atom density matrix evolves in time according to the
Markovian master equation given
by~\cite{lehmberg70:_radiat_n,agarwal74:_quant_statis_theor_spont_emiss_relat_approac,ficek02:_entan}
\begin{eqnarray}
  \label{eq:1}
 \frac{\partial \rho}{\partial t} &=&
  -i\sum_{i=1}^{2}\omega_{i}
  \left[S^{z}_{i},\rho\right]
  -i\sum_{i\neq j}^{2}{\Omega_{ij}}
  \left[S^{+}_{i}S^{-}_{j},\rho\right]\nonumber  \\
  &&- \frac{1}{2}\sum_{i,j=1}^{2}{
    \Gamma_{ij}}\left(\rho 
    S_{i}^{+}S_{j}^{-}+S_{i}^{+}S_{j}^{-}\rho
    -2S_{j}^{-}\rho S_{i}^{+}\right)
\end{eqnarray}
where $S_{i}^{+}$ ($S_{i}^{-}$) are the raising (lowering) operators,
and $S^{z}_{i}$ is the energy operator of the $i$th atom,
$\Gamma_{ii}\equiv \Gamma$ are the spontaneous decay rates.
We assume that the two
atoms are identical.  The parameters $\Gamma_{ij}$ and $\Omega_{ij}\
(i\neq j)$ depend on the distance between the atoms and describe the
collective damping and the dipole-dipole interaction defined,
respectively, by
\begin{eqnarray}
  \Gamma_{ij} = \frac{3}{2}\Gamma\left( \frac{\sin  kr_{ij}}{kr_{ij}}
    +\frac{\cos  kr_{ij}}{\left( kr_{ij}\right) ^{2}}-\frac{\sin kr_{ij} }
    {\left( kr_{ij}\right) ^{3}}\right)  ,\label{e2}
\end{eqnarray}
and
\begin{eqnarray}
  \Omega_{ij} = \frac{3}{4}\Gamma\left(-\frac{\cos  kr_{ij}}{kr_{ij}}
    +\frac{\sin kr_{ij} }{\left( kr_{ij}\right)^{2}}+\frac{\cos kr_{ij} }
    {\left( kr_{ij}\right) ^{3}}\right) ,\label{e3}
\end{eqnarray}
where $k =\omega_{0}/c$, and $r_{ij}$ is the distance between the
atoms. Here, we assume, with no loss of generality, that the atomic
dipole moments are parallel to each other and are polarised in the
direction perpendicular to the interatomic axis.

To describe the evolution of the two-qubit system the standard basis
of atomic product states can be used:
$|1\rangle=|e_{A}\rangle\otimes|e_{B}\rangle$,
$|2\rangle=|e_{A}\rangle\otimes|g_{B}\rangle$,
$|3\rangle=|g_{A}\rangle\otimes|e_{B}\rangle$,
$|4\rangle=|g_{A}\rangle\otimes|g_{B}\rangle$.  
It is easier, however, to find the solutions of the master equations
when using instead of the standard basis, a basis of the collective states:
$|e\rangle=|e_{A}\rangle\otimes|e_{B}\rangle$, 
$|s\rangle=\frac{1}{\sqrt{2}}\left(|e_{A}\rangle\otimes|g_{B}\rangle
  +|g_{B}\rangle\otimes|e_{B}\rangle\right)$, 
$|a\rangle=\frac{1}{\sqrt{2}}\left(|e_{A}\rangle\otimes|g_{B}\rangle
  -|g_{A}\rangle\otimes|e_{B}\rangle\right)$, 
$|g\rangle = |g_{A}\rangle\otimes|g_{B}\rangle$. 
The states $|s\rangle$
and $|a\rangle$ are the symmetric and antisymmetric states of the
two-atom system. They are maximally entangled states, or Bell states
of the two-atom system. 
Assuming that initially
the system density matrix has the so called $X$ form, which is
preserved during the evolution according to the master
equation~(\ref{eq:1}), and we get the following system of
equations for the density matrix elements~\cite{ficek02:_entan}
\begin{eqnarray}
  \label{eq:3}
  \rho_{ee}(t)&=&\rho_{ee}(0){\rm e}^{-2\Gamma t},\nonumber\\
  \rho_{ss}(t)&=&\rho_{ss}(0){\rm
    e}^{-\Gamma_{+}t}+\rho_{ee}(0)\frac{\Gamma_{+}}{\Gamma_{-}}\left[{\rm
      e}^{-\Gamma_{+}t}-{\rm e}^{-2\Gamma t}\right],\nonumber\\
  \rho_{aa}(t)&=&\rho_{aa}(0){\rm
    e}^{-\Gamma_{-}t}+\rho_{ee}(0)\frac{\Gamma_{-}}{\Gamma_{+}}\left[{\rm
      e}^{-\Gamma_{-}t}-{\rm e}^{-2\Gamma t}\right],\\
  \rho_{as}(t)&=&\rho_{as}(0){\rm e}^{-(\Gamma+i2\Omega_{12})t},\nonumber\\
  \rho_{eg}(t)&=&\rho_{ge}(0){\rm e}^{-(\Gamma+2i\omega_{0})t},\nonumber\
\end{eqnarray}
where $\Gamma_{\pm}=\Gamma\pm\Gamma_{AB}$.
For calculating the quantum correlations measures we need the
solutions for the density matrix elements in the standard product
basis, which can be expressed in terms of the matrix
elements~(\ref{eq:3}) in the following way
\begin{eqnarray}
  \label{eq:3a}
  \rho_{11}(t)&=&\rho_{ee}(t),\nonumber\\
  \rho_{14}(t)&=&\rho_{eg}(t),\nonumber\\
  \rho_{22}(t)&=&[\rho_{ss}(t)+\rho_{aa}(t)+\rho_{as}(t)+\rho_{sa}(t)]/2,\\
  \rho_{33}(t)&=&[\rho_{ss}(t)+\rho_{aa}(t)+\rho_{as}(t)-\rho_{sa}(t)]/2,\nonumber\\
  \rho_{23}(t)&=&[\rho_{ss}(t)-\rho_{aa}(t)+\rho_{as}(t)-\rho_{sa}(t)]/2.\nonumber
\end{eqnarray}
The solutions~(\ref{eq:3}) and~(\ref{eq:3a}) are used to find the
evolution of various measures of quantum correlations. 
\begin{figure}[t]
  \centering
  (a)\resizebox{0.45\textwidth}{!}{\includegraphics{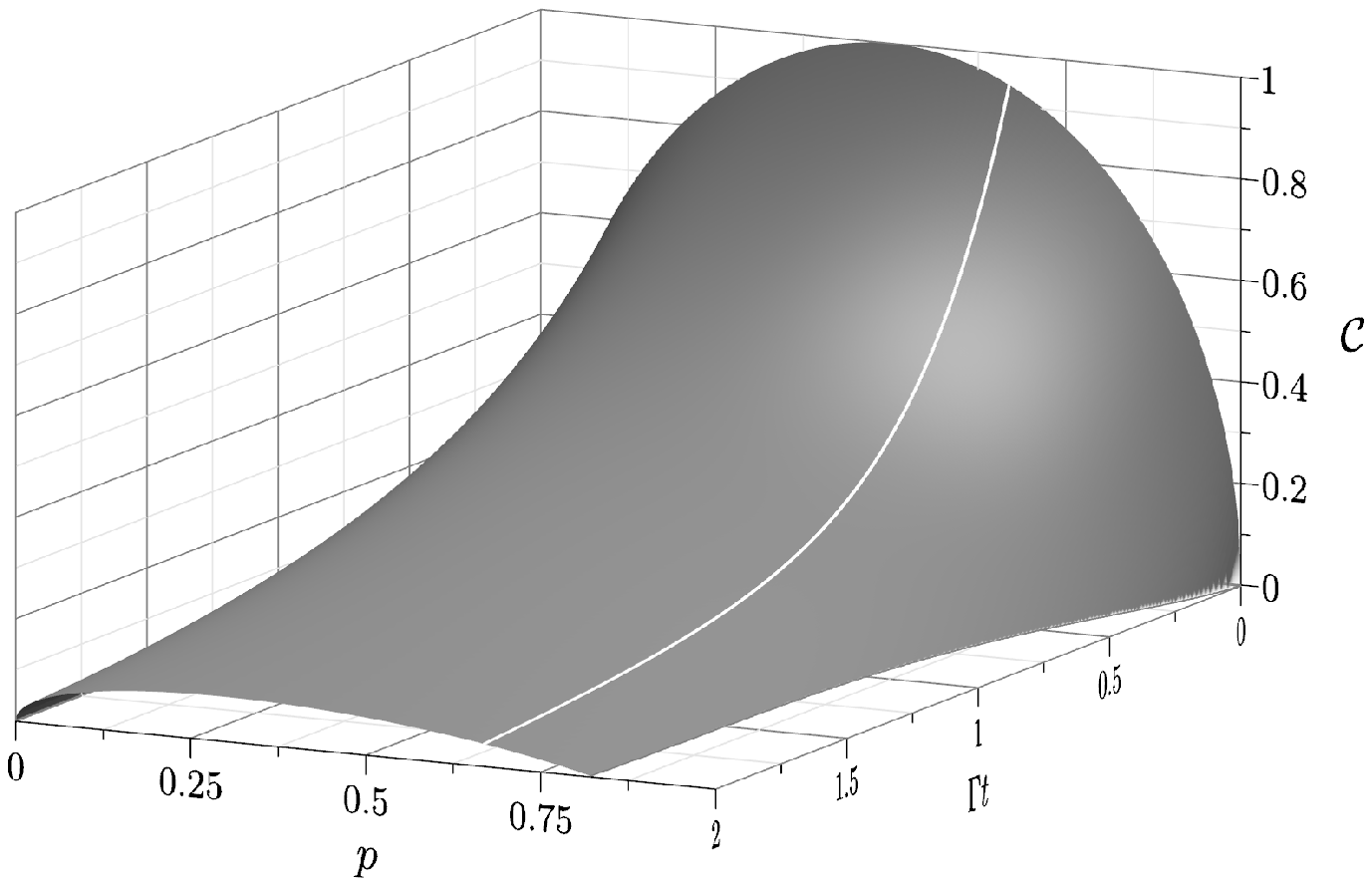}}\\[4pt]
  (b)\resizebox{0.45\textwidth}{!}{\includegraphics{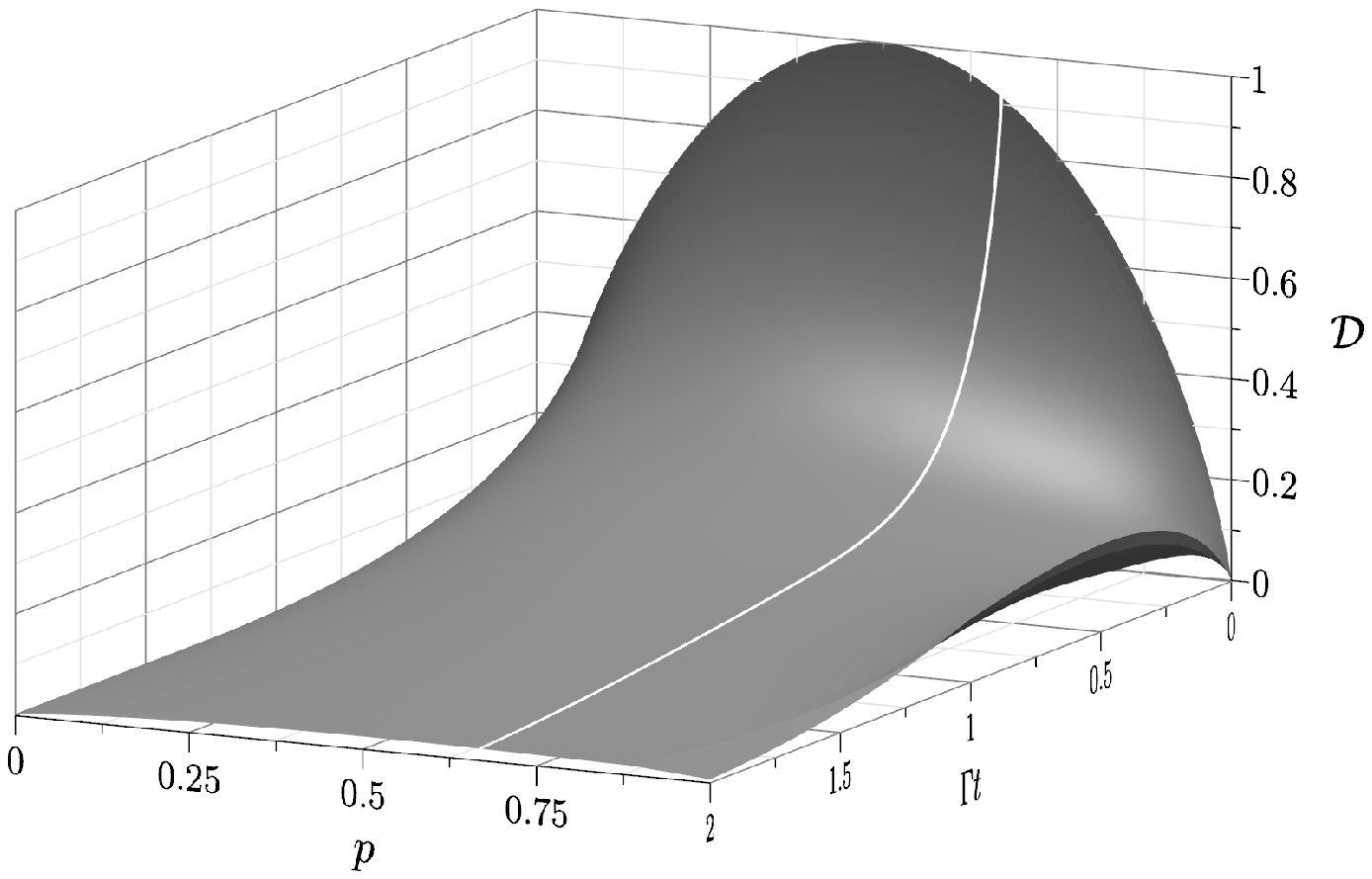}}\\[4pt]
  (c)\resizebox{0.45\textwidth}{!}{\includegraphics{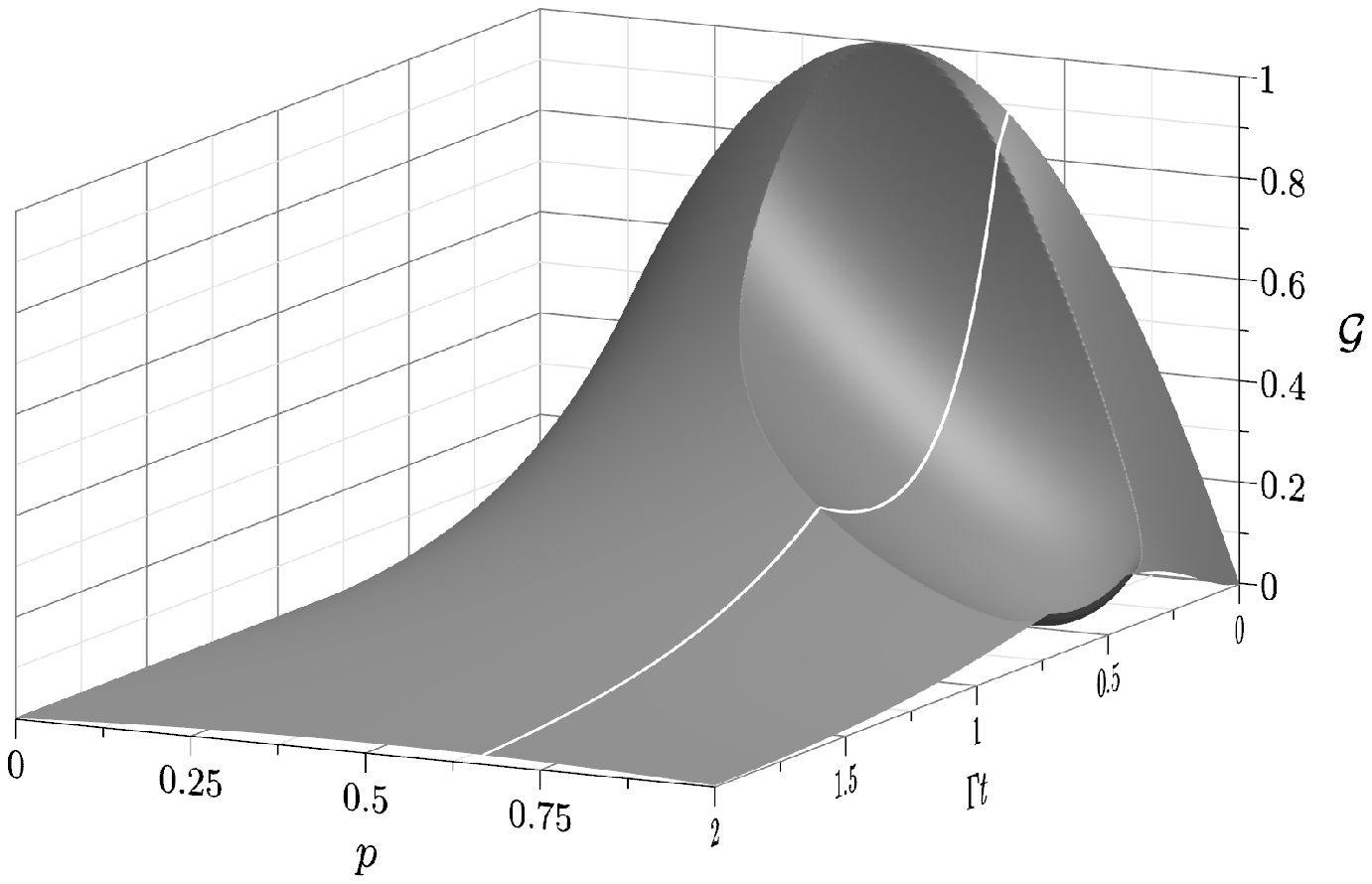}}
  \caption{Evolution of concurrence (a), quantum discord
    (b), and geometric quantum discord (c), for the two-photon
    Bell-like state~(\ref{eq:5}) and the interatomic distance
    $r_{AB}=\lambda/8$. The white lines show section of the surface for $p=2/3$.}
  \label{fig:1}
\end{figure}

\section{Measures of quantum correlations}
\subsection{Entanglement}
Probably the most celebrated and studied manifestation of quantum
correlations is quantum entanglement. 
To quantify the entanglement, various entanglement measures have been
introduced.  We use here concurrence introduced by
Wootters~\cite{wootters98:_entan}.
In the case of X states we consider, the
concurrence can be calculated analytically, and it has the
form~\cite{ficek03:_entan}
\begin{eqnarray}
  \label{eq:4}
  {\cal C}(t)&=&\max\left\{0,{\sf C}_{1}(t),{\sf C}_{2}(t)\right\}\nonumber\\
  {\sf C}_{1}(t)&=&2\left(|\rho_{14}(t)|-\sqrt{\rho_{22}(t)\rho_{33}(t)}\right)\\
  {\sf C}_{2}(t)&=&2\left(|\rho_{23}(t)|-\sqrt{\rho_{11}(t)\rho_{44}(t)}\right)\nonumber
\end{eqnarray}
Inserting into~(\ref{eq:4}) the solutions~(\ref{eq:3})
and~(\ref{eq:3a}), we find the values of ${\sf C}_{1}(t)$ and 
${\sf C}_{2}(t)$, and whenever one of the two quantities becomes
positive, there is a some degree of entanglement in the system.
\begin{figure}[t]
  \centering
  (a)\resizebox{0.45\textwidth}{!}{\includegraphics{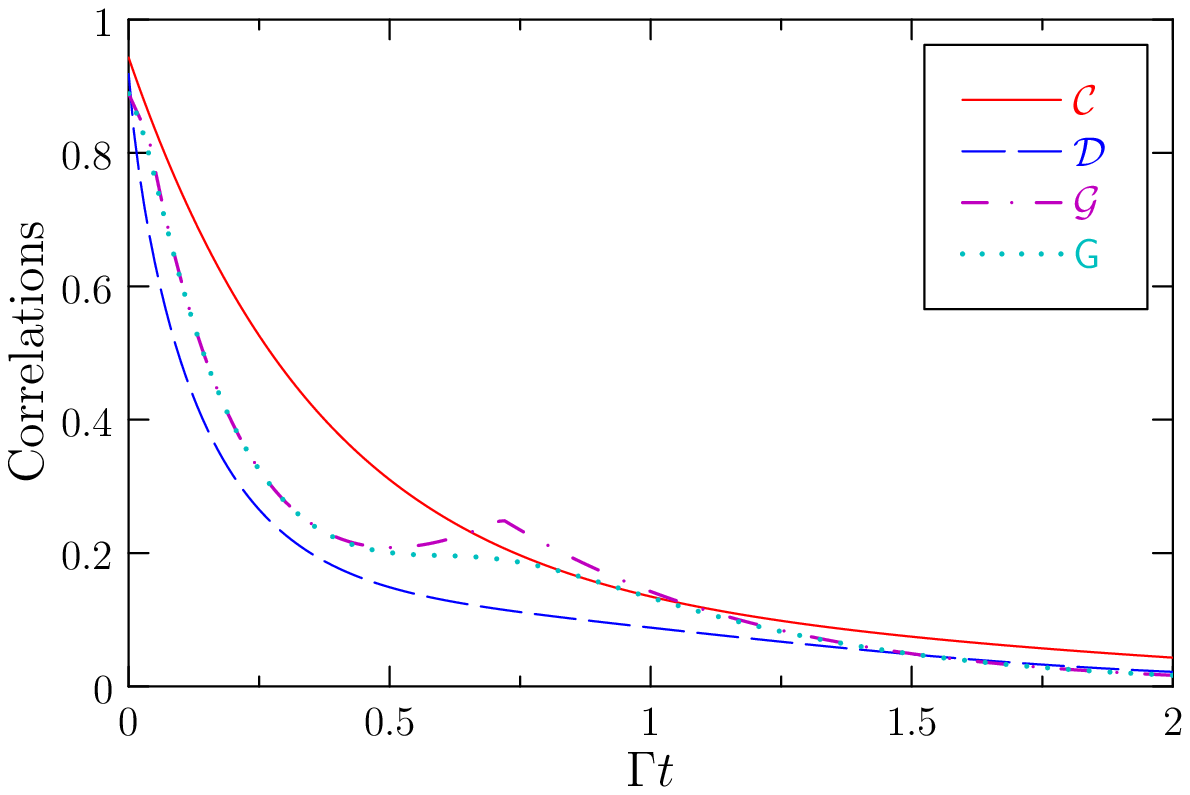}}\\[4pt]
  (b)\resizebox{0.45\textwidth}{!}{\includegraphics[width=7cm]{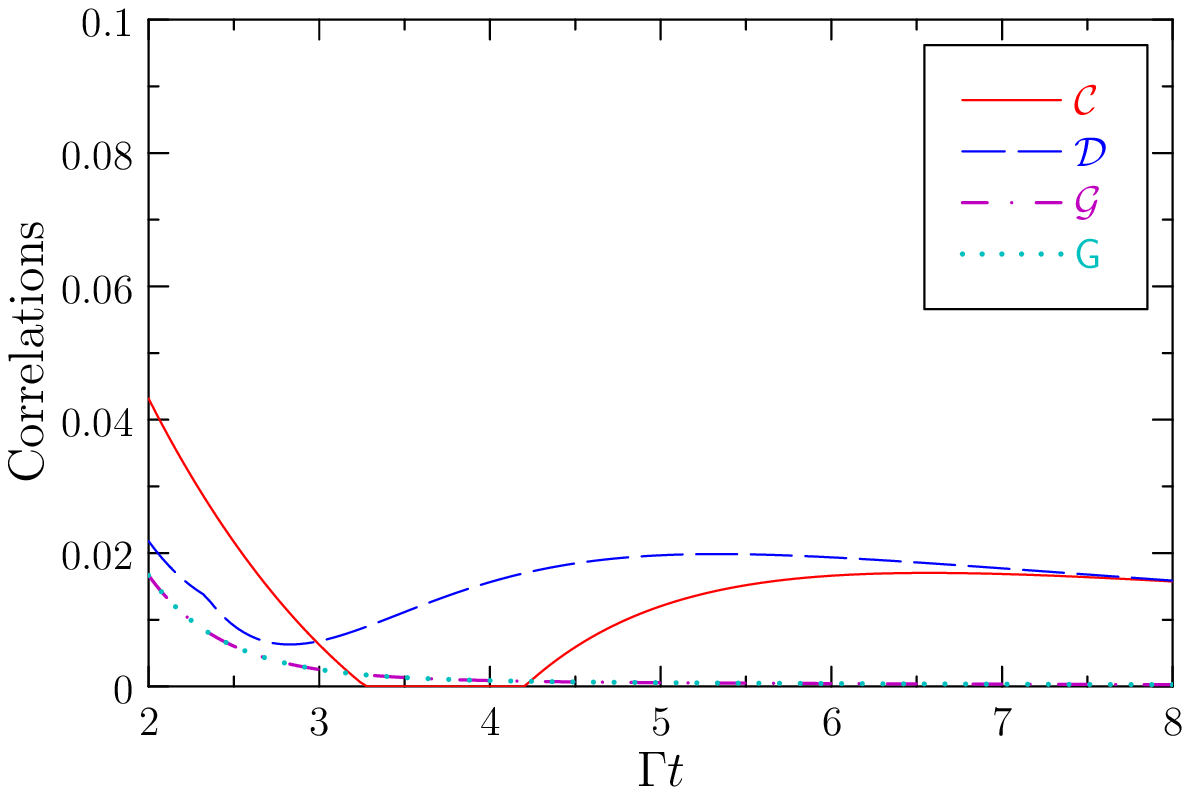}}
  \caption{Evolution of correlations for the
    state~(\ref{eq:5}) and interatomic distance $r_{AB}=\lambda/8$ for
    $p=2/3$. The evolution is split into two time regions (notice the
    different scales of both parts).}
  \label{fig:2}
\end{figure}
\begin{figure}[t]
  \centering
  (a)\resizebox{0.45\textwidth}{!}{\includegraphics[width=7cm]{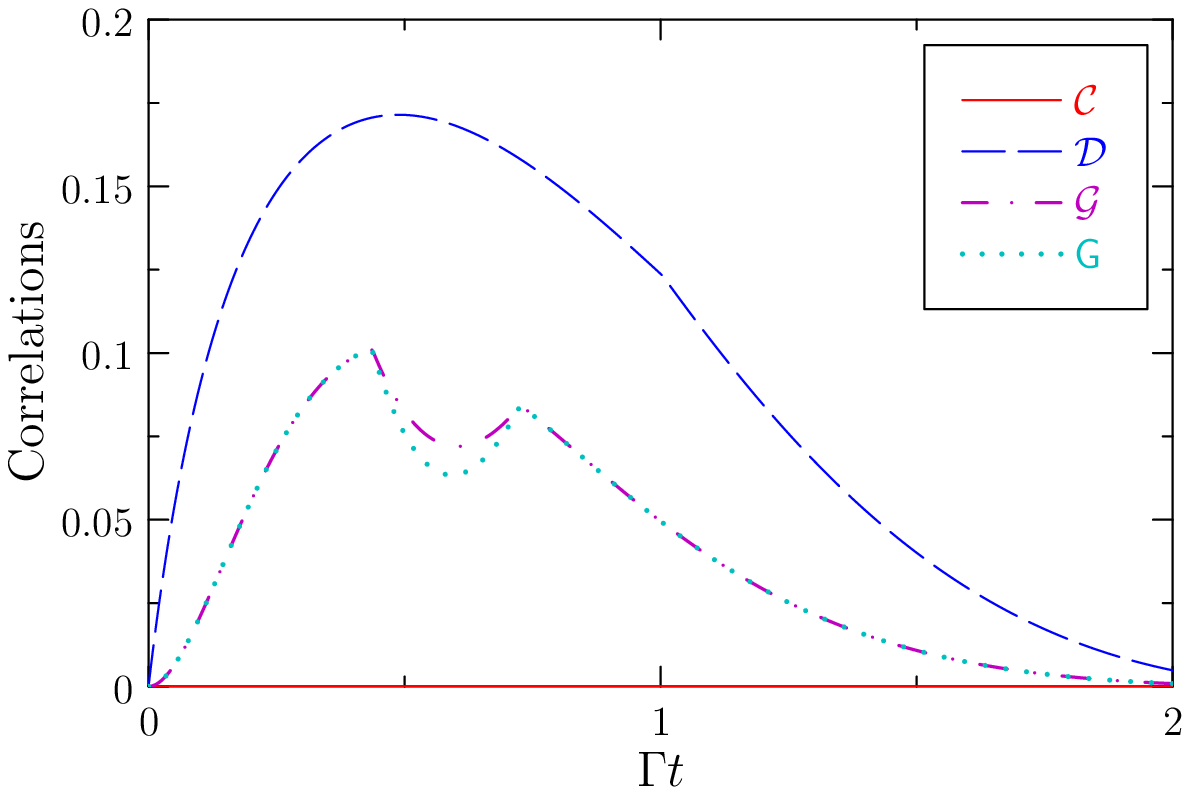}}\\[4pt]
  (b)\resizebox{0.45\textwidth}{!}{\includegraphics[width=7cm]{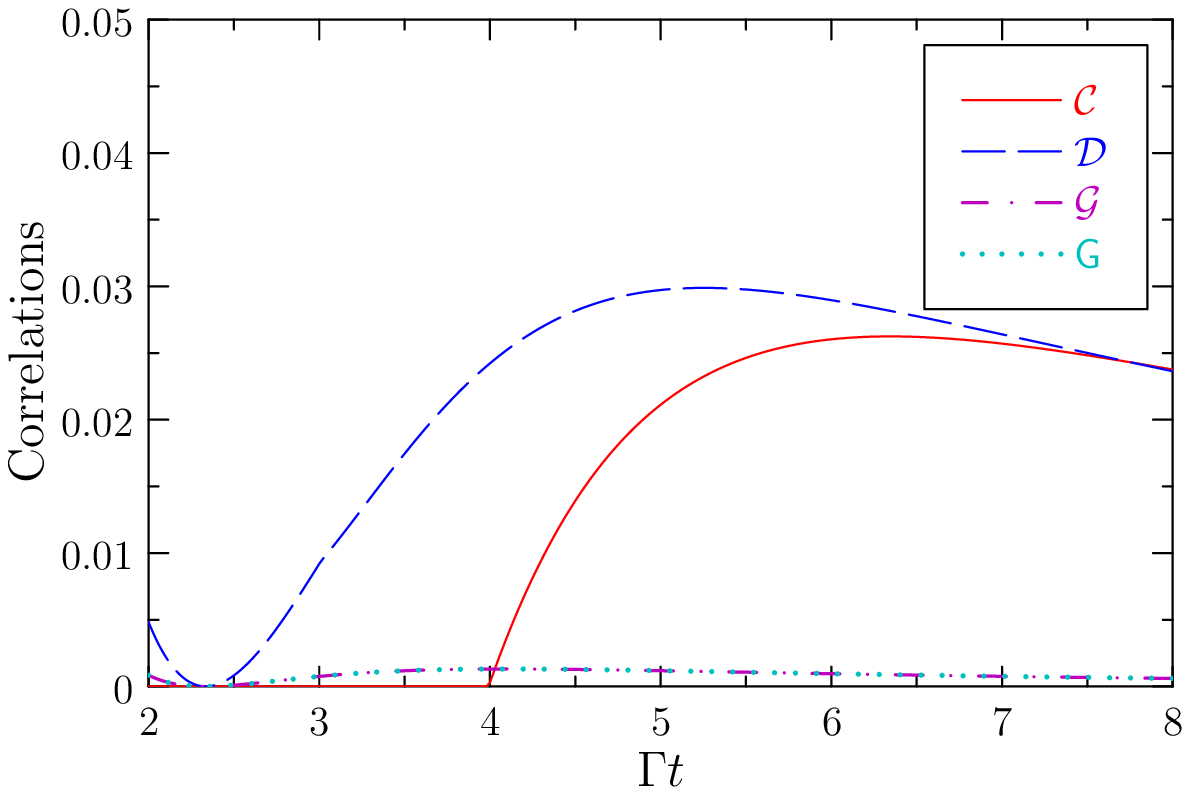}}
  \caption{Same as Fig.~\ref{fig:2}, but for $p=1$, {\em i.e.}
    initially both atoms excited.}
  \label{fig:3}
\end{figure}

\subsection{Quantum discord}
To calculate quantum discord we use the algorithm introduced by
Ali~{\em et al.}~\cite{ali10:_quant_x}, which works quite well in our
case. We have checked for many numerically generated X states, for
which the algorithm fails to find correct extremum, that the differences
between the numerically found extrema and the values given by the
algorithm are so small that they cannot be resolved in the scale of
the figures. In our calculation, we assume that the measurement is
performed on the subsystem $B$. For X state we
have~\cite{ollivier02:_quant,henderson01:_class,ali10:_quant_x}
\begin{eqnarray}
  \label{eq:10}
  {\cal D}&=&S(\rho^{B})-S(\rho)+\min\{{\cal K}_{1},{\cal
    K}_{2}\},\nonumber\\
  {\cal
    K}_{1}&=&H\left[\frac{1}{2}\left(1+\sqrt{(s^{A}_{z})^{2}+4\eta^{2}}\right)\right],\\
  {\cal K}_{2}&=&-\sum_{i}\rho_{ii}\log_{2}\rho_{ii}-S(\rho^{B}),\nonumber
\end{eqnarray}
where $S(\rho)$, $S(\rho^{B})$ mean the von Neumann entropy for the
two-atom system and the subsystem $B$, respectively,
$s_{z}^{A}={\rm Tr}[\rho\sigma^{A}_{z}]=\rho_{11}+\rho_{22}-\rho_{33}-\rho_{44}$ 
is the third component of the Bloch vector of the subsystem $A$, and
$\eta=|\rho_{14}|+|\rho_{23}|$. $H(x)$ is the Shannon entropy.

\subsection{Geometric quantum discord}
Geometric quantum discord is defined as~\cite{dakic10:_neces,luo10:_geomet}
\begin{eqnarray}
  \label{eq:11}
  {\cal G}&=&2\min_{{\cal X}}||\rho-{\cal X}||^{2}
\end{eqnarray}
where ${\cal X}$ is a set of zero-discord states, and $||\dots||$
denotes the Hilbert-Schmidt norm. The factor 2 in front has been
introduced for normalisation. The general formula for the geometric
quantum discord takes the form
\begin{eqnarray}
  \label{eq:13}
  {\cal G}&=&\frac{1}{2}\left(||\vec{s}^{B}||^{2}+||T||^{2}-k_{max}\right)
\end{eqnarray}
where $\vec{s}^{B}$ is the Bloch vector for the subsystem $B$ (we
assume that measurement is performed on subsystem $B$), and $k_{max}$
is the largest eigenvalue of the matrix
\begin{eqnarray}
  \label{eq:14}
K=\vec{s}^{B}(\vec{s}^{B})^{t}+T^{t}T,  
\end{eqnarray}
where the superscript $t$ means transposition.

For X states we have
\begin{eqnarray}
  \label{eq:12}
  {\cal G}&=&\min\{{\cal G}_{1},{\cal G}_{2}\}\nonumber\\
  {\cal G}_{1}&=&4(|\rho_{14}|^{2}+|\rho_{23}|^{2})\\
  {\cal G}_{2}&=&2(|\rho_{14}|^{2}-|\rho_{2}|^{2})^{2}+\frac{1}{2}\left[(s_{z}^{B})^{2}+T_{zz}^{2}\right]\nonumber
\end{eqnarray}
where $s^{B}_{z}=\rho_{11}-\rho_{22}+\rho_{33}-\rho_{44}$ is the third
component of the Bloch vector for the subsystem $B$ and $T_{ij}={\rm
  Tr}[\rho(\sigma^{A}_{i}\otimes\sigma_{j}^{B})$, ($i,j=x,y,z$),  are
elements of the correlation matrix.

The observable measure of Girolami and Adesso~\cite{girolami12:_obser}
is given by the formula
\begin{eqnarray}
  \label{eq:15}
  {\sl G}&=&\frac{1}{6}\left[2{\rm Tr}(K)-\sqrt{6{\rm
        tr}(K^{2})-2\left[{\rm Tr}(K)\right]^{2}}\right]
\end{eqnarray}
with the matrix $K$ given by~(\ref{eq:14}).

\section{Evolution of quantum correlations}
To compare the evolution of various measures of quantum correlations
in a system of two two-level atoms governed by the master equation~(\ref{eq:1}),
we assume that the initial state is the Bell-like state 
\begin{eqnarray}
  \label{eq:5}
|\Psi\rangle=\sqrt{p}\,|1\rangle+\sqrt{1-p}\,|4\rangle
\end{eqnarray}
where $p$ is the population of the upper state $|1\rangle$.
In Fig.~\ref{fig:1} we illustrate the evolution of concurrence,
quantum discord and geometric quantum discord for the whole range of
values of $p$. For $p<1$ the state~(\ref{eq:5}) is a superposition of
the states $|1\rangle$ and $|4\rangle$ which has nonzero two-photon
coherence $\rho_{14}$, while for $p=1$ it is a product state of both
atoms excited. It is seen from Fig.~\ref{fig:1} that the evolution of 
different measures of quantum correlations depends on the value of $p$
and shows essential differences, not only quantitative but also
qualitative. The white line seen on the figures indicates the section
of the surface at $p=2/3$. The evolution for this value of $p$ is illustrated in
more details in Fig.~\ref{fig:2}. It has been shown
before~\cite{ficek06:_dark} that there is sudden death
and revival of entanglement in the system for $p>0.5$, which is
clearly seen in Fig.~\ref{fig:2}b. The other correlations do not
exhibit sudden death, but there is an increase in quantum discord for
$\Gamma t>2$. The most striking feature is the behaviour of geometric quantum
discord with a deep crater for $\Gamma t<1$ as seen in Fig.~\ref{fig:1}c,
and the cusp seen in Fig.~\ref{fig:2}a. The observable measure of
geometric quantum discord given by~(\ref{eq:15}) and shown in
Fig.~\ref{fig:2}a smoothes out the cusp of geometric discord. 

For $p=1$ the state~(\ref{eq:5}) becomes the product state of both
atoms being excited for which all the correlations are zero. However, as it
is evident from Fig.~\ref{fig:1}, concurrence remains zero for some
time, while the other correlations increase immediately after the
start of the evolution. It is clearly evident
from Fig.~\ref{fig:3}. For short times the quantum
discord increases, reaches the maximum and goes down to the minimum in
order to increase again to the subsequent maximum, and eventually it
goes down to zero asymptotically. This behaviour has been reported
in~\cite{auyuanet10:_quant}. The concurrence remains zero up to
$\Gamma t\sim 4$ and next abruptly becomes nonzero, the effect that
has been referred to as sudden birth of
entanglement~\cite{ficek08:_delay}. Both revival of entanglement and
the sudden birth of entanglement are due to collective behaviour of
the two atoms when the interatomic distance is smaller than the
wavelength of light emitted by individual
atom~\cite{tanas04:_entan,ficek06:_dark,ficek08:_delay}. Here we take
the interatomic distance equal to $\lambda/8$. 

From Fig.~\ref{fig:3}a it is seen that the geometric quantum discord
has a minimum with sharp cusps, and the observable measure of
geometric quantum discord, which is a lower bound for geometric quantum
discord, is a really tight bound, except for the interval around the
minimum, where the bound is not so tight. For this time interval
geometric quantum discord behaves quite differently than quantum
discord, and both differ from concurrence.   

\section{Conclusions}
We have discussed the dynamics of various measures of quantum correlations
in a two-atom system
interacting with a common reservoir in a vacuum state. The
evolution of the system is described by the Lehmberg-Agarwal Markovian
master equation, which takes into account collective behaviour of the
atoms. The collective spontaneous emission is a source of quantum
correlations in the system. We have shown that the evolution of
different measures of quantum correlations is qualitatively different,
with a rather strange behaviour of the geometric discord. Some aspects
of the evolution of quantum correlation in such a system has
been studied in~\cite{hu12:_robus}. Recently Piani~\cite{piani12} has
argued that the geometric discord is not a good measure for the
quantumness of correlations.

\section*{Acknowledgement}
I thank Zbyszek Ficek for fruitful discussions on the subject. 
This work was supported by the Polish National Science
Centre under grant 2011/03/B/ST2/01903. 

\section*{References}

\providecommand{\newblock}{}

\end{document}